\documentclass[11pt]{iopart}

\usepackage{iopams}
\usepackage{graphicx}

\begin{document}
\marginpar{\texttt{compiled: \today}}
\title[WD MLDC1]{Inference on white dwarf binary systems using the first round Mock LISA Data Challenges data sets}
 \author{Alexander~Stroeer${^{1,2}}$, John~Veitch$^1$, Christian~R\"over$^3$, 
         Ed~Bloomer$^4$,
         James~Clark$^4$, Nelson~Christensen$^5$,
         Martin~Hendry$^4$, Chris~Messenger$^4$,
         Renate~Meyer$^3$, Matthew~Pitkin$^4$,
         Jennifer~Toher$^4$,
         Richard~Umst\"atter$^3$, 
         Alberto~Vecchio$^{1,2}$ and Graham~Woan$^4$}
 \address{$^1$ School of Physics \& Astronomy,
          University of Birmingham, Birmingham, UK}
 \address{$^2$ Department of Physics \& Astronomy,
          Northwestern University, Evanston, IL, USA}
 \address{$^3$ Department of Statistics, The University of Auckland,
          Auckland, New Zealand}
\address{$^4$ Department of Physics \& Astronomy,
          University of Glasgow, Glasgow, UK}
 \address{$^5$ Physics \& Astronomy,
          Carleton College, Northfield, MN, USA}

\begin{abstract}
We report on the analysis of selected single source data sets from the first
round of the Mock LISA Data Challenges (MLDC) for white dwarf binaries. 
We implemented an end-to-end pipeline consisting of a grid-based coherent pre-processing 
unit for signal detection, and an automatic Markov Chain Monte Carlo post-processing unit for signal evaluation. We demonstrate that signal detection with our coherent approach is secure and accurate, and is increased in accuracy and supplemented with additional information on the signal parameters by our Markov Chain Monte Carlo approach. We also demonstrate that the Markov Chain Monte Carlo routine is additionally able to determine accurately the noise level in the frequency window of interest. 

\end{abstract}


\submitto{Classical and Quantum Gravity}

\section{Introduction}
\label{s:intro}

The data obtained from LISA~\cite{Benderetal:98} will contain a large number of white dwarf binary systems across the whole observational window~\cite{Nelemansetal:2001}. At frequencies below $\sim 3$ mHz the sources are so abundant that they produce a stochastic foreground whose intensity dominates the instrumental noise~\cite{FarmerPhinney:2003}. The closer (and louder) sources will still be sufficiently bright to be individually resolvable. Above $\sim 3$ mHz the sources become sufficiently sparse in parameter space (and in particular in the frequency domain) that the detectable sources become individually resolvable. The identification of white dwarfs in the LISA data set represents one of the most interesting analysis problems posed by the mission: the total number of signals in the data set is unknown, the effective noise level affecting the measurements is not easily estimated from the data streams, and there is a large number of overlapping sources to the limit of confusion.

Bayesian inference provides a clear framework to tackle such a problem~\cite{Jaynes:2003,Gregory:2005,Gelmanetal:1997}. Some of us have carried out exploratory studies and ``proof of concept'' analyses on simplified problems that have demonstrated that Bayesian techniques do indeed show good potential for LISA applications~\cite{Umstaetter:2005,Stroeeretal:2006,Wickhametal:2006}. Similarly other authors have successfully implemented techniques using Bayesian inference~\cite{CornishCrowder:2005,Crowderetal:2006,CrowderCornish:2007}. In this paper we present the first results of an end-to-end analysis pipeline developed in the context of the Mock LISA Data Challenges that has evolved from our earlier work. This pipeline is applied to the simplest single-source challenge data sets 1.1.1a and 1.1.1b and all the results presented here are obtained \emph{after} the release of the key files. In a companion paper~\cite{Roeveretal:2007}, we present results that we have obtained for the analysis of the data sets containing gravitational radiation from a massive-black-hole binary inspiral. Our group submitted an entry for the MLDC analysing the blind data set 1.1.1c~\cite{glig-report1,mldc1-summary}: however that result suffered from the fact that the pipeline was not complete, the analysis code was inefficient and we encountered hardware problems with the Beowulf cluster used to perform the analysis. 

The results that we present here are obtained with a two-stage end-to-end analysis pipeline: (i) we first process the data set with a grid-based coherent algorithm to identify candidate signals; (ii) we then follow up the candidate signals with a Markov Chain Monte Carlo code to obtain probability density function on the model parameters. Our method differs from other MCMC methods that have been proposed and applied to the MLDC data in the context of white dwarf binaries~\cite{CornishCrowder:2005,Crowderetal:2006,CrowderCornish:2007}: the MCMC is not used to search, but only in the final stage of the analysis to produce posterior density functions of the model parameters. The noise spectral level is included as one of the unknown parameters and is estimated together with the parameters of the gravitational wave source(s). 

\section{Analysis method}

In this section we describe the two stage approach that we have adopted for the analysis. The signal produced by a white dwarf binary system is modelled as monochromatic in the source reference frame, following the conventions adopted in the first MLDC~\cite{MLDCLISA06a,MLDCLISA06b,MLDCdoc}. It is described by 7 parameters: ecliptic latitude $\vartheta_e$ and longitude $\varphi_e$, inclination $\iota$ and polarisation angle $\Psi$, frequency at a reference time $f_0$ and corresponding overall phase $\Phi_0$ and  amplitude $A$. 

The data distributed for the MLDC are the three TDI v1.5 Michelson observables $X$, $Y$ and $Z$~\footnote{In our MCMC analysis we use the data set produced using the LISA Simulator.}. From those we construct the two orthogonal TDI outputs
\begin{eqnarray}
      A &=& (2X-Y-Z)/3 \\ 
      E &=& (Z-Y)/\sqrt{3}
\label{e:AE}
\end{eqnarray}
by diagonalizing the noise covariance matrix following the procedure presented in~\cite{Princeetal:2002}. The noise affecting the channels $A$ and $E$ is uncorrelated and described by the one-sided noise spectral density $S_n(f)$. We model the LISA response function in the low frequency limit in order to improve the computational efficiency of our analysis.

\subsection{First stage: Grid based search}

The first stage of the pipeline consists of a fast search of the data for the best matched filter based on the well-known $\mathcal{F}$-statistic algorithm, first developed for triaxial pulsar signals in the context of ground-based observations~\cite{JKS}. This exploits the Fast Fourier Transform to perform matching in the frequency domain to templates which are generated at an array of fixed points in the parameter space. 

The data from an individual detector in the frequency domain $\tilde{d}(f)$ is supposed to contain a signal plus Gaussian noise, $\tilde{d}(f) = \tilde{h}(f) + \tilde{n}(f)$. We define the logarithmic likelihood as a measure of match, as given by $\log{}L\approx (\tilde{d}|\tilde{h}) - \frac{1}{2}(\tilde{h}|\tilde{h})$ with $(\cdot|\cdot)$ denoting the scalar product as defined in \cite{JKS}.
A single signal in the $\mathcal{F}$-statistic algorithm is re-parameterised as a linear function of four orthogonal variables, and the frequency $f_0$. The detection statistic is based on four parameters $A_{\mathcal{F}}$, $B_{\mathcal{F}}$, $C_{\mathcal{F}}$ and $D_{\mathcal{F}}$, found by integrating over the response functions $a(t)$ and $b(t)$ to the two polarisation states of the gravitational wave signal \cite{JKS}, 
\begin{eqnarray}
A_{\mathcal{F}} & = & \frac{2}{T_{\rm obs}}\int_0^{T_{\rm obs}}a(t)^2dt \\
B_{\mathcal{F}} & = & \frac{2}{T_{\rm obs}}\int_0^{T_{\rm obs}}b(t)^2dt \\
C_{\mathcal{F}} & = & \frac{2}{T_{\rm obs}}\int_0^{T_{\rm obs}}a(t)b(t)dt, \\
D_{\mathcal{F}} & = & A_{\mathcal{F}}B_{\mathcal{F}}-C_{\mathcal{F}}^2
\end{eqnarray}
$T_{\rm obs}$ denotes the total observed time for the data set being analysed.
The optimal detection statistic $2\mathcal{F}$, which is pre-maximised over the nuisance parameters $h_0$, $\iota$, $\phi_0$ and $\psi$ is
\begin{equation}
2\mathcal{F}=\frac{8}{S_n(f)T_{\rm obs}}\frac{B_{\mathcal{F}}|F_a|^2+A_{\mathcal{F}}|F_b|^2-2C_{\mathcal{F}}\times{}{\mathcal{R}}(F_aF_b)}{D_{\mathcal{F}}}.
\end{equation} 
$F_a$ and $F_b$ are the demodulated Fourier transforms of the data,
\begin{equation}
F_a=\int_0^{T_{\rm obs}} d(t)a(t)e^{-i\Phi(t)}dt; \,\,\,F_b=\int_0^{T_{\rm obs}} d(t)b(t)e^{-i\Phi(t)}dt,
\end{equation}
$\Phi(t)$ is the phase of the gravitational wave signal, as is described in \cite{Brady1997}.

\begin{figure}
\resizebox{0.9\hsize}{!}{\includegraphics{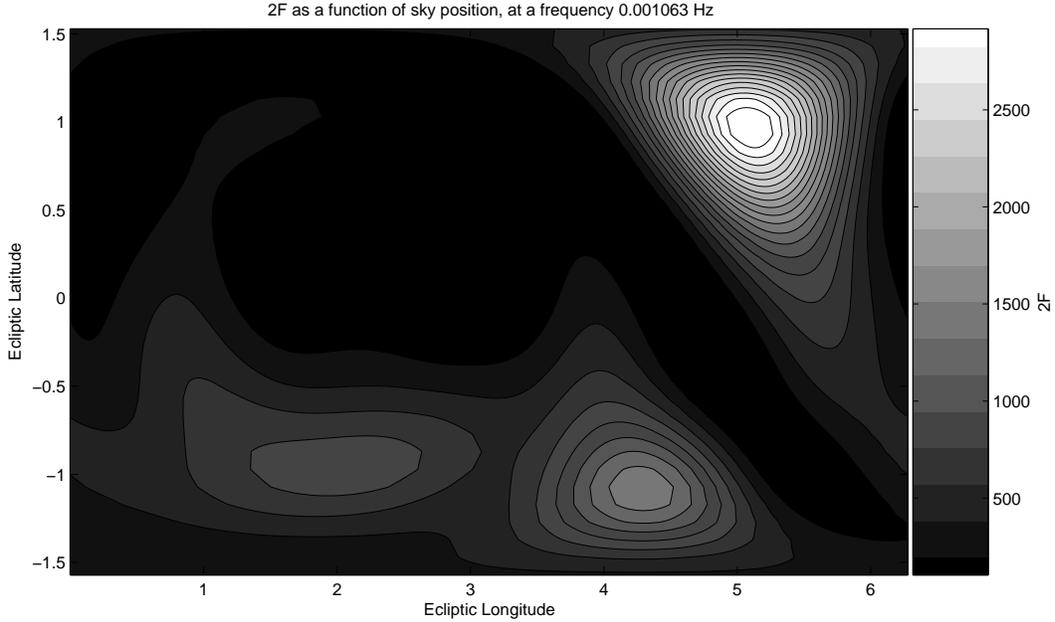}}
\caption{\label{f:skypos} The variation of $2\mathcal{F}$ values for the search for unknown signal 1.1.1a, as a function of sky position, parameterised by ecliptic latitude $\beta$ and longitude $\lambda$. The distribution is multi-modal and non-Gaussian, and has a poor resolution in comparison with that can be achieved with the MCMC and a Bayesian likelihood, but by finding the maximum it serves well as a starting point for the more refined parameter estimation below.}
\end{figure}

As the LISA array moves in space, the frequency $f_0$ is affected by Doppler modulations. This modulation changes with differing position of the source in the sky, implying the need to recalculate the modulations and thus $a(t)$ and $b(t)$ for each sky position that is tested - a significant factor in the performance of this approach. The differing modulation structure however also allows us to estimate the location of the source in the sky by maximising the $2\mathcal{F}$ value. The resolution possible on the sky with this method is not as good as from a full Bayesian posterior probability calculation as performed in the parameter estimation stage, as shown in an example for Challenge 1.1.1a in figure \ref{f:skypos}. Nevertheless, since this statistic can be computed fairly quickly it serves as a useful way of finding initial values to feed into the MCMC routine, as adopted within the pipeline. The resolution achievable on the sky increases with frequency, which implies that the mismatch between filter and signal falls off more rapidly at higher frequencies, requiring a greater number of templates to cover the sky. Therefore for challenge 1.1.1b at $f\approx3\,{\rm mHz}$ a sky grid of size 5,752 points was used, in comparison with 765 points for challenge 1.1.1a at $f\approx1\,{\rm mHz}$.

The $\mathcal{F}$-statistic search was implemented using the LIGO ``Lalapps'' suite of software \cite{LAL}, in which the pulsar search code was modified by Reinhard Prix and John Whelan to use the LISA response function for the TDI variables $X$, $Y$, and $Z$ \cite{PrixWhelan}. These input data streams were given in the form of Short Fourier Transforms, each of length one day, created from the MLDC1 challenge data. For each challenge the full specified range of frequencies was searched for the signal as it would be in a blind search.
The code was run on a single CPU and executed in a few hours, with the run-time increasing at higher frequency due to the higher resolution of sky and frequency grid that had to be used. The candidate chosen to pass to the MCMC stage was simply that which triggered the highest value of $2\mathcal{F}$. 

\subsection{Second stage: Markov Chain Monte Carlo follow-up}
According to Bayes' theorem, the posterior probability, $p(\tilde{m}|\tilde{d})$ of a model $\tilde{m}$ given the data $\tilde{d}$ depends on the prior distribution $p(\tilde{m})$, containing the information known before the analysis, the likelihood $L(\tilde{d}|\tilde{m})$ of the model and a normalisation factor $p(\tilde{d})$
\begin{equation}
p(\tilde{m}|\tilde{d})=\frac{L(\tilde{d}|\tilde{m})p(\tilde{m})}{p(\tilde{d})}
\end{equation}
The posterior probability density function shows the joint probability density of given values of parameters of the model $\tilde{m}$, conditional on the data $\tilde{d}$. 

We implemented Bayes' theorem using data in the form of TDI variables $A$ and $E$ and modelled our template according to the Long Wavelength Approximation directly in the Fourier domain \cite{CornishLarson:2003} to gain computational speed. 
The logarithmic likelihood $L(\tilde{d}|\tilde{m})$ in this stage explicitly included its dependence on the one-sided noise spectral density $S_n(f)$ 
\begin{equation}\label{likelihood}
\log{L(\tilde{d}|\tilde{m})} = {\rm const.} \; - \; \frac{1}{2}\log{S_n(f)} \; - \;  (\tilde{d}-\tilde{h}|\tilde{d}-\tilde{h}) ,
\end{equation}
shown here for either $A$ or $E$, with the combined likelihood as sum of the individual likelihoods.
We restricted our analysis to a sufficiently narrow frequency window in order to be able to approximate the noise spectral density as constant, $S_n(f)=S_0$. This window was set as the interval in frequency that contains at least 98\% of the power of our model $\tilde{m}$, with the interval's upper and lower limits given by  $f\pm (2/{\rm year})(5 + 2\pi f_0 AU/c)$ \cite{CornishLarson:2003}. $S_0$ is therefore an additional parameter to be inferred within the model $\tilde{m}$ in Eq.~\ref{likelihood}.

We implemented an automatic Random Walk Metropolis sampler (Stroeer \& Vecchio 2007, in. prep.) to sample from the posterior probability density function in form of a Markov chain. Metropolis sampling eliminates the need to explicitly calculate the normalisation constant in Bayes' theorem, and the evolving Markov chain gives easy access to joint as well as marginalised posterior density distribution. The sampler was started from the parameter set which triggered the highest value of $2\mathcal{F}$ in our grid based coherent run of the analysis (see former section). The automated function of the Metropolis sampling is achieved by controlling the sampling step-size with adaptive acceptance probability techniques \cite{AtchadeRosenthal:2005}. The sampler therefore does not depend on assumptions about the signal in the data set in order to perform successfully and reliably; it develops a suitable algorithm and approach by itself based on the properties of the likelihood as found on the fly, in the initial steps of the sampler. The length of our Markov chain was pre-set to $10^6$, with the initial $10^4$ chain states discarded as the ``burn-in'' phase of our sampler. The runtime for one data analysis run is 5 hours on a single 2 GHz CPU on the Tsunami cluster of the University of Birmingham. 

\section{Results}

\begin{figure}
\resizebox{\hsize}{!}{\includegraphics{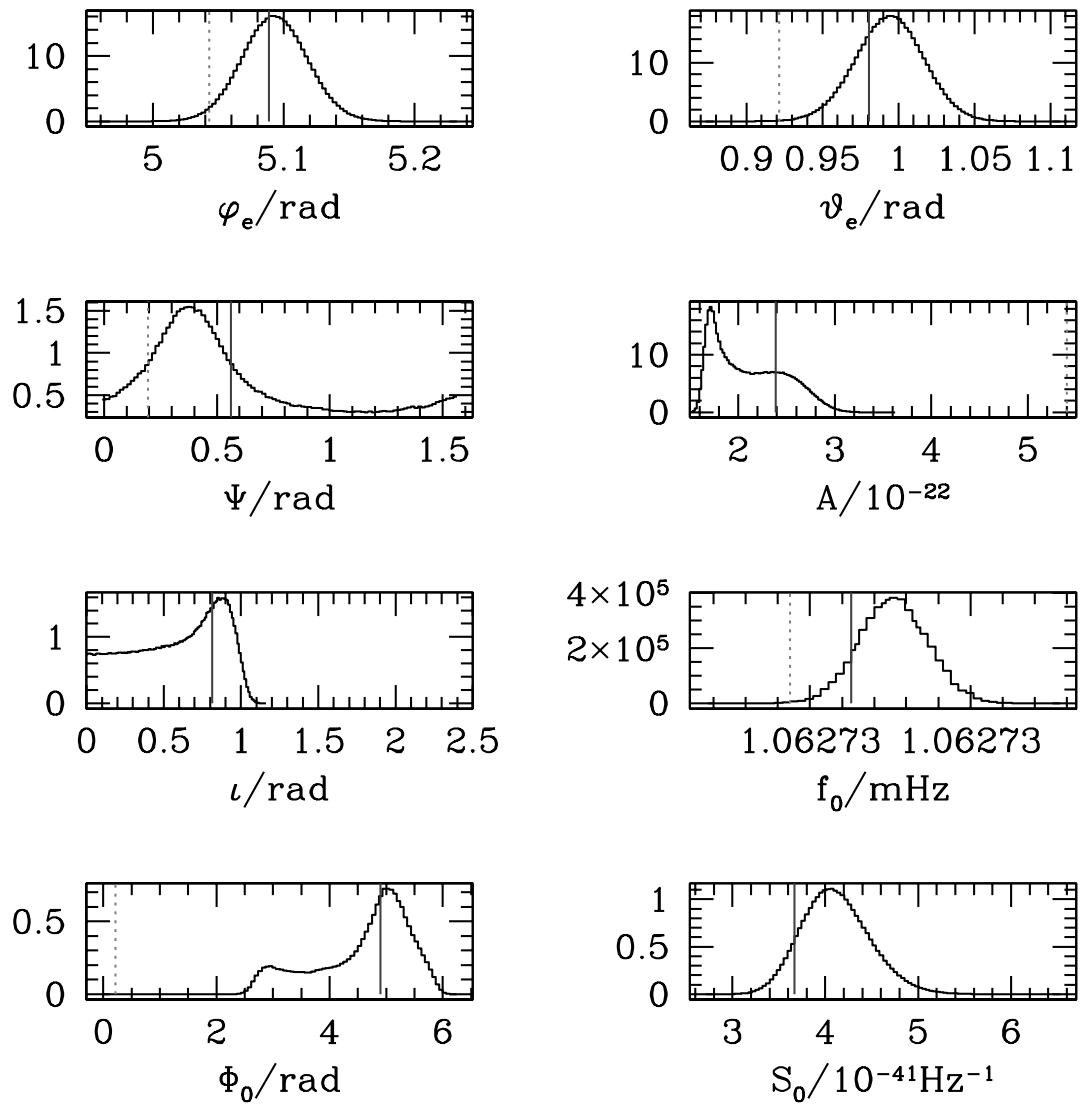}}
\caption{The marginalised posterior probability density functions of the eight unknown parameters -- the seven parameters that describe the signal and the noise spectral density $S_0$ -- for the the challenge data set 1.1.1a. The vertical black solid line denotes the true value of the parameter (for the polarisation angle the true value modulo $\pi/2$), and the grey dashed line the initial value for the MCMC analysis as determined by the template of the first-stage that produces the maximum value of the ${\cal F}$-statistic. In the case of the noise spectral density the first stage of the analysis does not provide an estimate; the true value of this parameter is taken to be the value of the instrumental noise spectrum used to generate the data set and provided in~\protect{\cite{MLDCdoc}}.}
\label{f:111a}
\end{figure}

\begin{figure}
\resizebox{\hsize}{!}{\includegraphics{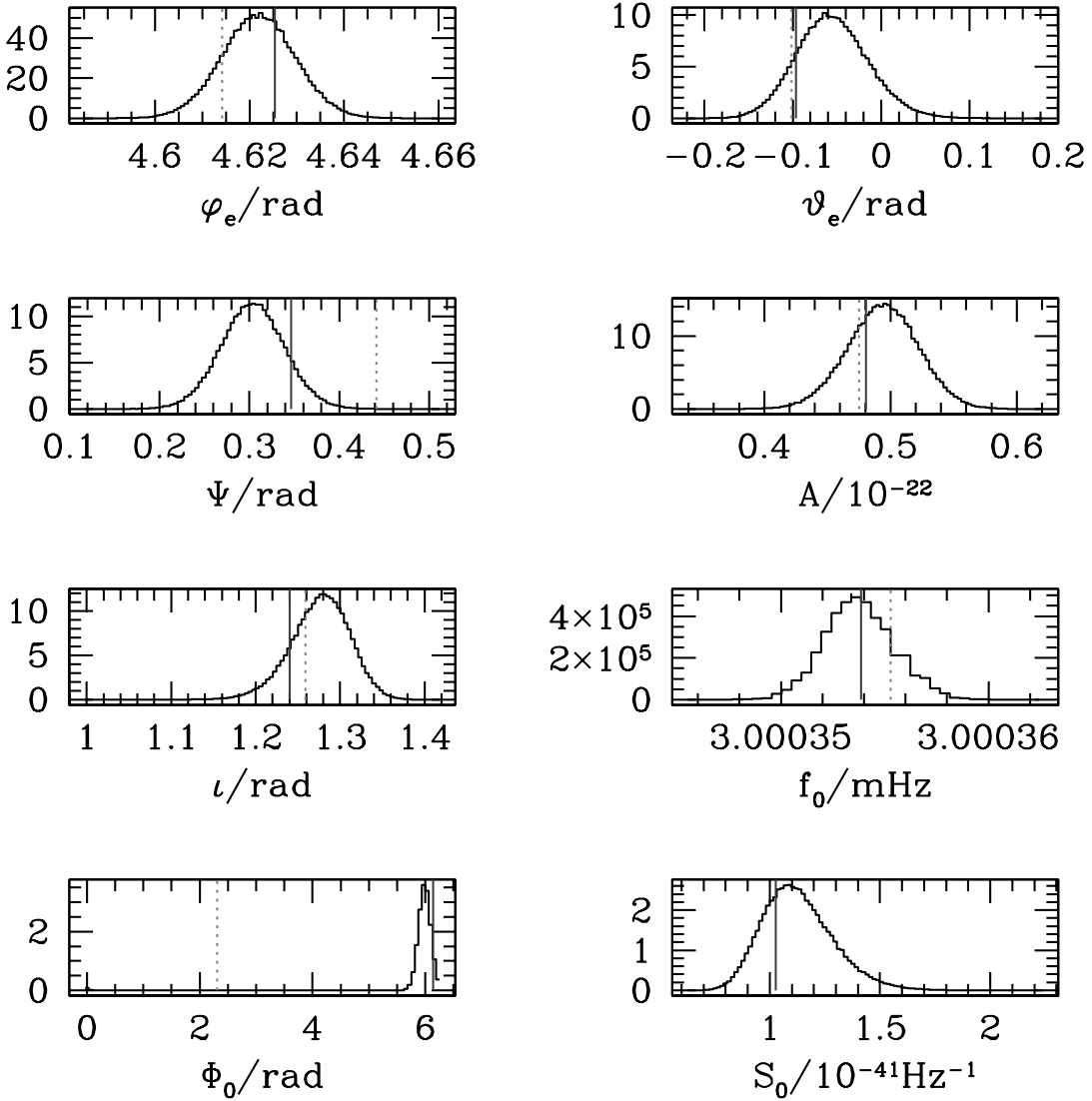}}
\caption{The marginalised posterior probability density functions of the eight unknown parameters for the the challenge data set 1.1.1b. Labels are as in Figure 2.}
\label{f:111b}
\end{figure}

\begin{table}
\caption{\label{t:111}Details about the results from Challenge 1.1.1a and Challenge 1.1.1b. $S_0$, the constant one-sided noise spectral density within our narrow frequency window, is compared to the true one sided noise spectral density at the true frequency of the signal, $\Psi$ is given  modulo $\pi/2$. Int$_{90}$  denotes the minimum interval to include 90$\%$ of MCMC states for given parameter, $\Delta$mode denotes the absolute difference between the true value of a signal parameter and the mode of its recovered posterior; $\Delta$median and $\Delta$mean denote the equivalent absolute difference for median and mean of the posterior respectively; $\sigma$ denotes the sampled standard deviation of the posterior as derived from the median. We further quote the signal-to-noise ratio (SNR) for a template using the true values of the source and the recovered values of the data analysis run, as derived from the median of the individual posterior distributions, and the correlation C between these two templates.}
\small{
\begin{tabular}{l|r|r|r|r|r}
\hline
&Int$_{90}$ & $\Delta$mode & $\Delta$median & $\Delta$mean & $\sigma$ \\ 
\hline
\hline
\multicolumn{6}{l}{Challenge 1.1.1a}\\
\hline
\hline
$\frac{S_0}{10^{-41}Hz^{-1}}$  &(3.53257, 4.72639) & -0.42084 & -0.440278 & -0.452456 & 0.36704  \\               
$\vartheta_e / $ rad           &(0.958409, 1.03165) & -0.0147383 & -0.0149381 & -0.0148725 & 0.0222861 \\             
$\varphi_e / $ rad             &(5.05376, 5.13528) & -0.00550139 & -0.00569547 & -0.00579889 & 0.0247886 \\       
$\Psi / $ rad                  &(1.32475, 0.500553) & 0.1768 & 0.1823 & 0.1902 & 0.1908 \\                 
$\iota / $ rad                 &(0.097761, 1.0008) & -0.0459747 & 0.190001 & 0.23459 & 0.295211 \\                
$A/10^{-22}$                   &(1.61976, 2.67967) & 0.664371 & 0.358844 & 0.298978 & 0.368524 \\                 
$f_0 / $ mHz                   &(1.06273, 1.06273) & -1.19664e-06 & -1.22207e-06 & -1.22259e-06 & 1.04422e-06 \\  
$\Phi_0 / $ rad                &(3.10668, 5.808) & -0.164989 & 0.00998525 & 0.229659 & 0.829146 \\                                        
\hline
SNR&\multicolumn{2}{c|}{true = 51.024497}&\multicolumn{3}{c}{recovered = 50.648600}\\
\hline
C&\multicolumn{5}{c}{true vs. recovered = 0.99689}\\
\hline
\hline
\multicolumn{6}{l}{Challenge 1.1.1b}\\
\hline
\hline
$\frac{S_0}{10^{-41}Hz^{-1}}$   &(0.876833, 1.38959) & -0.0679571 & -0.0906557 & -0.0996144 & 0.16017  \\     
$\vartheta_e / $ rad            &(-0.121611, 0.0116916) & -0.0343353 & -0.151185 & -0.150328 & 0.0406552 \\   
$\varphi_e / $ rad              &(4.60969, 4.63537) & 0.00265723 & 0.00305564 & 0.00302203 & 0.00779893 \\    
$\Psi / $ rad                   &(0.246328, 0.362409) & 0.0301541 & 0.0311747 & 0.0311268 & 0.0353938 \\      
$\iota / $ rad                  &(1.22036, 1.33338) & -0.0430412 & -0.040458 & -0.0394818 & 0.0348383 \\      
$A/10^{-22}$                    &(0.45001, 0.542454) & -0.016442 & -0.0151921 & -0.0149907 & 0.0281154 \\     
$f_0 /$ mHz                     &(3.00036, 3.00036) & 3.1221e-07 & 2.49289e-07 & 2.42807e-07 & 8.18111e-07 \\ 
$\Phi_0 / $ rad                 &(5.83869, 6.19411) & 0.137219 & 0.119301 & 0.119921 & 0.502384 \\            
\hline
SNR&\multicolumn{2}{c|}{true = 36.587444}&\multicolumn{3}{c}{recovered = 37.368806}\\
\hline
C&\multicolumn{5}{c}{true vs. recovered = 0.97897}\\
\hline
\hline
\end{tabular}}
\end{table}

We found that the most promising candidate signal from the $\mathcal{F}$-statistic search already matched the true embedded signal to high accuracy, particularly in frequency and sky location. Our MCMC sampler, as a post-processing unit, thus only needed 1000 iterations to burn in and to establish a reliable sampling from the posterior. The marginalised posteriors are shown in Figs.~\ref{f:111a} and ~\ref{f:111b}. We found, as seen in latter figures, that the MCMC sampler further refined the initial guesses from the $\mathcal{F}$-statistic, as measured by the absolute difference between the true value of a given parameter and the median of the marginalised posterior recovered for that parameter. Table~\ref{t:111} shows details of the statistics of recovered posterior distributions. We highlight that the majority of the true values of the parameters are within one standard deviation of the median of the posterior, with a small percentage within two sampled standard deviations. In addition, every true value of a parameter of the signal is within the minimum interval of the posterior to cover 90$\%$ of all MCMC state values. Recovered signal-to-noise ratios are measured as ${\rm SNR} = {\left(s|h\right)}/{\sqrt{\left(h|h\right)}}$, and the match $C = {\left(h_{\rm true}|h_{\rm med}\right)}/{\sqrt{\left(h_{\rm true}|h_{\rm true}\right)\left(h_{\rm med}|h_{\rm med}\right)}}$ between a template constructed from the true values and a template from the median values of the individual posterior distributions, yielding a correlation that is always higher than 0.97. Noise levels are determined accurately and within 1 to 1.5 sampled standard deviations. Nevertheless we note that our run on Challenge 1.1.1a shows a lower match and higher differences between true value and recovered value of parameters as compared to the run on Challenge 1.1.1b. It also exhibits tailing posterior distributions in inclination and amplitude, although the SNR of Challenge 1.1.1a is twice the value of Challenge 1.1.1b.

\section{Conclusions}
We have presented a new approach to LISA data analysis in the form of an end-to-end pipeline. We first detected and identified candidate signals in the LISA data stream with a grid-based coherent algorithm, and then post-processed the most promising candidate signals with an automatic Markov Chain Monte Carlo code to obtain probability densities for the model's parameters. We demonstrated successful identification and post-processing of the signals from the double white dwarf single source MLDC data sets 1.1.1a and 1.1.1b. Furthermore, the automatic Markov Chain Monte Carlo code successfully identified the noise level within a small frequency window of interest in these data sets. We note that a parallel approach to the data analysis of binary inspiral signals is being developed by R\"{o}ver \emph{et al}, with a Markov Chain Monte Carlo method that can successfully post-process a candidate signal generated from the true parameters of the signal. Signal detection in a pre-processing stage is currently being tested within parallel tempered MCMC methods and/or time-frequency analyses \cite{Roeveretal:2007}.

We identify two prominent and promising features of our pipeline: its ability to determine good initial conditions for the MCMC and its ability to run the MCMC automatically. As we have demonstrated in this paper, the width of the marginalised posterior density for the frequency parameter is extremely narrow. It is therefore vital that the initial estimate of the frequency is within this region, as the almost flat structure of the posterior PDF outside this region gives little to no information on the location of the peak. The chances of finding the mode through a random sampling are decreased further still with a larger prior range for the parameter. Adding an $\mathcal{F}$-statistic search as the first stage in the pipeline solves this problem, since the frequency and position in the sky are recovered very accurately, to within the limits of the posterior probability region of interest, before the MCMC performs post-processing and parameter estimation. The automatic feature of the MCMC ensures a successful post-processing for the other astrophysical parameters that may have been located outside the posterior region of interest by the $\mathcal{F}$-statistic approach, as in the case for the amplitude of Challenge 1.1.1a. Convergence is aided by the ability of our code to increase or decrease sampling step-sizes according to its experience of the sampling quality of the posterior during the burn-in phase.

We are working on an extension of the pipeline as shown in this document to successfully tackle multi-source data sets, required for the second round of the MLDC. Current work includes the exploration of our grid-based coherent search on such data streams in order to automatically identify the most promising individual candidate signals, and the implementation of an automatic Reversible Jump Markov Chain Monte Carlo routine (e.g. as already demonstrated in \cite{Stroeeretal:2006}) to find the trans-dimensional probability density functions of the parameters of an unknown total number of signals. We highlight that the noise level determination presented here already serves as a key ingredient to round 2, where the simulation of a galactic white dwarf binary population introduces additional confusion noise levels from unresolvable sources.
\section*{Acknowledgements}
Nelson Christensen's work was supported by the National Science Foundation grant PHY-0553422 and the Fulbright Scholar 
Program. Alberto Vecchio's work was partially supported by the Packard Foundation and the National Science Foundation. The 
University of Auckland group was supported by the Royal Society of New 
Zealand Marsden Fund Grant UOA-204.

\end{document}